# WIRELESS IP TELEPHONY

Mohsen Gerami

The Faculty of Applied Science of Post and Communications
Danesh Blv, Jenah Ave, Azadi Sqr, Tehran, Iran.
Postal code: 1391637111

*Abstract*—**The convergence of traditional telecommunications and the Internet is creating new network-based service delivery opportunities for telecommunications companies carriers, service providers, and network equipment providers. Voice over Wireless IP is one of the most exciting new developments emerging within the telephony market. It is set to revolutionize the delivery of mobile voice Services and provide exciting new opportunities for operators and service providers alike. This survey discusses principal of Wireless IP Telephony.**

 Keywords- IP telephony; Convergence; Wireless; Internet; Wi-Fi;

## I. INTRODUCTION

"IP will eat everything" meaning all systems and networks will eventually use Internet-based protocols. Convergence of communications and applications will become a reality-- networks will be the computer. There is a huge computing power sitting on company networks. Wireless Internet will be big and will drive mobility. [1]

Wireless VoIP utilizes wireless LAN technology, the same wireless infrastructure used for your corporate network, in order to communicate. Just as you use PDAs and laptops to gain access to information within this wireless infrastructure, now you can use wireless IP phones to access your corporate telephony system as this technology combines the telephony function directly into an already existing data network infrastructure.

One of the major benefits of the wireless IP phone is that it allows you to carry your office extension with you inside a wireless networked environment. Unlike your cell phone, the wireless IP phone is part of your corporate phone system, and carries your personal extension and the same features that your office phone system has.

IP telephony offers many benefits to users in both large and small organizations, but the major gain will be in productivity. By extending mobile communications throughout the enterprise, wireless IP telephony helps users increase their productivity when they are not working at their desk. By enabling users to answer critical business calls anywhere anytime within a wireless campus environment, improved business response results.

Cost savings are also realized with a wireless IP phone system because it offers easy mobility for organizations where employee offices change often. Additionally, with VoIP telephony, expanding the communications system is easier and less costly. Because the wireless infrastructure is designed to handle voice and data, new employees can be assigned a phone and instantly be on line and mobile without having to install lines and jacks.

An estimated $7.6 billion will be spent on wireless data services by 2006. Industries that are ripe for this technology include healthcare, manufacturing, transportation, and education -- any industry where people need highly mobile, feature rich communications capabilities in a campus like environment.

A large cargo shipping container terminal company is preparing to deploy wireless IP phones at their terminal facilities. They already have an 802.11b wireless infrastructure in place to support a tablet PC-based manifest and custom application, so the addition of the wireless IP phones was a natural next step. Equipped with wireless IP phones, the cargo handlers and custom brokers can stay in constant communication no matter where they are in the container yard

Manufacturing -- Constant Communications Even in the Factory!

Employees walking the shop floors need access to their office phones. Today, they may be carrying 2-way radios or cell phones to communicate. With wireless IP phones, they can carry their real office phone with them wherever they want, no matter where they are within the facility.

As this technology becomes more available, you'll see it everywhere. Eventually you may be able to walk into a coffee shop, a Starbucks for example that's wired with this technology, and use your IP phone. Pay a small subscription fee and upon entering the coffee shop your phone activates itself, connects back to your corporate network, and gives you access to your office extension while sitting there. The possibilities are very real and are here today. [2]

## II. WIRELESS INTERNET TELEPHONY

As Figure 1 shows, a recent report by the Yankee Group, a market research firm, predicts the US consumer Internet telephony market will explode from 130,000 subscribers at the end of 2003 to 17.5 million subscribers in 2008.

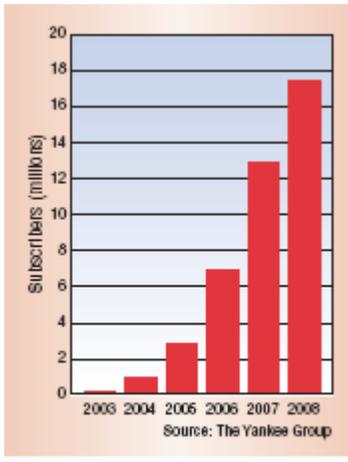

*Figure 1. The Yankee Group, a market research firm, predicts rapid growth in the US Internet telephony market.*

Now, providers are offering wireless Internet telephony, which adds convenience by letting users make Internet calls from their mobile phones via IEEE 802.11 (Wi-Fi) wireless LAN and thirdgeneration (3G) cellular technologies.

As is the case with its wired Internet telephony, wireless Internet telephony is less expensive than regular mobile telephony because carriers can use the existing Internet, rather than build a new infrastructure, to route calls. In addition, Internet telephony is not subject to the regulation and fees that governments impose on traditional telephony.

Internet telephony faces several important concerns, particularly power usage, security, and quality of service (QoS).

In many ways, wireless Internet telephony is an adaptation of traditional wireline IP telephony, as the "Internet Telephony 101" sidebar explains. Wireless IP telephony works primarily with Wi-Fi, which it uses to access the Internet. However, many

Internet calls do not travel only over Wi-Fi networks. For example, a call from a user on a Wi-Fi network to someone using a traditional wireline or mobile phone at some point will be routed over the traditional wired or cellular phone network. Some systems provide wireless service only via Wi-Fi. For example, Spectra Link's system connects its Master

Control Unit to Wi-Fi base stations on one end and to a traditional analog or digital PBX on the other, thereby eliminating the need for cellular service.

However, several companies, including Motorola, are developing phones that would use cellular technology for the parts of calls that travel over cellular networks and Wi-Fi for those parts that travel over the Internet.

### A. Wi-Fi

In Wi-Fi Internet telephony, vendors equip a mobile handset with an IEEE 802.11 radio. The phones, when within range of a Wi-Fi access point, use IEEE 802.11 to connect to the Internet, over which they can then transmit voice traffic.

There are several Wi-Fi standards. IEEE 802.11b, the first popular Wi-Fi standard, has a theoretical maximum data rate of 11 Mbits per second using the 2.4-GHz frequency band. IEEE 802.11a has a theoretical maximum rate of 54 Mbps using the 5-GHz band. IEEE 802.11g offers a faster speed and compatiblity with the large installed base of IEEE 802.11b systems because it also uses the 2.4-GHz band. Wi-Fi works with telephony by providing a wireless channel to the Internet. Wi-Fi converts voice and other data into radio signals that can be transmitted wirelessly. Internet-connected receivers then convert the radio signals into conventional data traffic that can be transmitted via the Internet or another network.

There are a growing number of Wi-Fi-enabled networks and IEEE 802.11 phones from manufacturers such as Cisco Systems and Symbol Technologies, said Allen Nogee, principal analyst for wireless technology with In-Stat/MDR, a market research firm. Companies such as Agere Systems, Broadcom, and Texas Instruments (TI) are beginning to release Wi-Fi-based Internet telephony chips, which have embedded functionality formerly provided by both software and hardware, said Allen.

For example, TI's TNETV1600 system- on-chip platform consists of a voice-over-IP application processor, an IEEE 802.11b and IEEE 802.11g media-access-control baseband processor, and a radio transceiver. Originally, Wi-Fi worked only within a wireless LAN. In recent years, though, individuals and companies have established Wi-Fi hot spots, which are nodes that provide laptops, cellular phones, and other mobile devices within the technology's range of 100 meters indoors and 400 meters outdoors with Internet connections. Many hot spots are close to one another, which gives cellular-phone users widespread Internet access. In-Stat/MDR predicts that sales of business-class Wi-Fi-based Internet phones will increase about 120 percent from 2003 to this year, while the Wi-Fi Internet telephony market will grow from $16.5 million in 2002 to $500 Million by 2007.

### B. Internet Telephony Concerns

Despite its promise, wireless Internet telephony raises several important concerns. For example, the relatively new Wi-Fi Internet phones can be quite expensive. However, Nogee noted, prices are dropping.

Also, Wi-Fi Internet phones use considerable power for their radio transmissions.

This requires bigger phones to accommodate more or larger batteries at a time when the demand is for smaller phones.

### C. Security

As with any wireless technology, Internet telephony raises security concerns. For example, authentication approaches, which determine whether people trying to access a system are who they say they are, must be improved and standardized, said Nogee.

Without standardization, competing technologies from vendors could cause system incompatibilities.

Meanwhile, wireless security itself is only now overcoming some early problems. Many users complained that the Wired Equivalent Privacy Protocol, used in early Wi-Fi applications, was not strong enough. The industry has since evolved to Wi-Fi Protected Access and now WPA2, based on the IEEE 802.11i wireless-security standard, which uses the new Advanced Encryption Standard. Regardless, wireless Internet telephony security will require the exchange of additional information between senders and receivers and thus slow data transmissions, which will affect voice quality, explained ON World's Hatler.

### D. Quality of service

QoS is a key issue for Internet telephony. IP networks must prioritize telephony traffic because, unlike other data traffic with which it shares the Internet, voice data must be transmitted in real time. If not, voice quality degrades and latency becomes a problem.

Currently, several approaches, mostly proprietary, provide some QoS for Internet telephony. According to Nogee, the proposed IEEE 802.11e standard would provide a way to address Wi-Fi-based QoS. The standard, currently under consideration, would accomplish this by prioritizing packets based on traffic type, enabling access points to schedule resources based on transmission rates and latencies, and otherwise improving bandwidth efficiency. [3]

### III. SIMPLE IDEA, COMPLEX EXECUTION

The technology behind voice over wireless — varyingly referred to as wireless IP telephony, wireless VoIP, and Wi-Fi telephony — is straightforward. Mobile handsets connect to the network over wireless access points, routing the voice traffic to the telephony server or digital PBX in the same way that VoIP handsets connect to the network over Ethernet cables, routing their voice traffic to the telephony server or digital PBX . That apparent simplicity is why many enterprises consider implementing voice over wireless when they implement a VoIP system. Chances are they're also deploying wireless access points for data usage, so they believe that most of the infrastructure required to make VoIP mobile is already in place.

☐ Wireless LANs for voice require denser access-point placement to reduce contention for the access point's bandwidth and require deployment in areas like hallways, elevator shafts, and facilities service areas in which data usage would not occur. That means greater hardware and installation costs.

☐ The number of simultaneous calls over an access point is limited to anywhere from four to two dozen, depending on the wireless LAN's implementation and architecture, as well as actual usage.

☐ Users can experience interruptions or even dropped calls due to contention for access points.

☐ Wireless handsets cost about $300 to $400 each.

☐ Wireless LAN deployments tend to occur before VoIP deployments, so enterprises must over engineer their wireless LANs to accommodate future VoIP deployments.[4]

### IV. CONVERGENCE

One of the greatest advantages of the New World IP telephony system is the ease of intelligent integration with existing applications. In the New World, IP PBX voice mail and e-mail are all part of the same application running in a distributed fashion across the entire corporate network. A single mailbox can now hold a user's voice messages, e-mail, fax, and video clips.

Convergence is being driven by cost and by applications that demand voice/data integration, such as real-time distance learning, videoconferencing, integrated voice mail and e-mail messaging, and voice-enabled desktop applications.

### A. Data/Voice/Video Integration Cost Effective

If you look at the overall bandwidth requirements of voice compared to the rest of the data network, then it is miniscule. On a per-packet or per-kilobit charge, voice is basically free. Therefore, adding voice to a data network is very cost effective.

### B. Operations Simplified

One of the greatest advantages of the integrated voice and data system is the ease of intelligent integration with existing applications.

End users can use their Web browsers to define graphically a personal rules engine that can filter incoming calls, scan and organize voice mail, create personal phone configurations such as speed dial, and build a valet service that scans a personal calendar to intelligently route calls. A single mailbox can now hold voice messages, e-mail, fax, and video clips.

Another benefit is that expensive PBX equipment can be eliminated. Traditional PBX call routing and embedded features are based on proprietary applications that are specific to that particular system. Traditional PBXs are like an island, independent of all the other applications running on the corporate network. In the new system, IP PBX, voice mail, and e-mail are all part of the same application that runs in a distributed fashion across the entire corporate network.

### C. Competitive Advantage

The Internet has created the capability for almost any computer system to communicate with any other. With Internet business solutions, companies can redefine how they share relevant information with the key constituents in their business, not just their internal functional groups, but also customers, partners, and suppliers.

This ubiquitous connectivity created by Internet business solutions creates tighter relationships across the company's extended enterprise, and can be as much of a competitive advantage for the company as its core products and services.[5]

## V.   HOW VOIP WORKS

VoIP (Voice over Internet Protocol) is the way to use the advantages of internet to transmit voice and place cheap international calls. VoIP technologies are designed to take analog audio signals like that used on traditional telephone lines and to turn them into digital data like that used on the Internet.

Why VoIP is so popular today? VoIP solutions are simple to use and unbelievably cheap. The reason is that VoIP uses standard internet connection to place phone calls therefore you do not need to use the services of phone companies. Hence you pay only for your internet connection. Besides, most VoIP providers offer either free international calls or low cost international calls and VoIP providers are likely to supply you with VoIP software.

A number of VoIP service providers have already been working in the market for a number of years and settled several calling plans so that you may choose the one that meets your needs best. VoIP services expand greatly and telecommunication companies are to offer more and more VoIP options for customers.

VoIP is a convenient replacement of a traditional phone system. What make VoIP more attractive is that there are several ways to use VoIP services today:

VoIP services are available through using your ordinary phone with the ATA (analog telephone adaptor). The ATA connects your phone line to the internet and convert the analog data of the telephone lines to digital data of the internet and vice versa. When you purchase VoIP services the ATA will very probably be included into the package supplied by VoIP providers. To make the entire system work you should only connect your phone to the ATA with the cable that usually goes to the socket. Some types of the ATA are released with VoIP software to configure the ATA on your computer. However the VoIP adapters are very simple to use and do not require special maintenance.

VoIP phones are used instead of ordinary phones to make calls using VoIP services. IP phones look like ordinary phones but have an Ethernet connector and work with digital data. IP phones are connected to the router and have all the hardware and VoIP software necessary for placing and receiving calls through IP.

The easiest way to perform VoIP communication is to place calls from computer to computer. VoIP software is offered by many VoIP companies for free and moreover you can make free long distance calls. To use VoIP services through your computer you should have Internet connection, a sound card, a microphone and speakers [6].

## VI.   REGULATORY ISSUE

Should VoIPbe regulated? Why? What form of regulation is appropriate?

Should some existing requirements of voice telephone services be abolished or changed?

Should there should be regulatory forbearance to allow VoIPto develop in the market?

What happens to telephone numbers?

How can universal service obligations, emergency call features, lawful access etc. be achieved in this environment?

Initial Responses to VoIP

Some regulators have removed restrictions; in developing countries, most regulators have applied restrictions

VoIP competition has reduced prices significantly

In developed countries, incumbent operators response is to bundle:

National tariffs, but excluding fixed to mobile

DSL plus telephony (video etc.)

Offering in-bound numbers in other countries

In developing countries, most incumbents have tried to restrict VoIP [7].

## VII.   THE DISAVANTAGES OF IP TELEPHANY

Despite their relatively young age VoIP technologies have already started to replace traditional phone systems. The number of customers that prefer IP telephony to other means of telecommunications grows rapidly that may lead to a complete dominance of IP telephony in the telecommunication market. Eventually phone companies and businesses are likely to switch to VoIP services completely.

However VoIP technologies as any other emerging technologies have a lot to think about and modernize. VoIP providers can't satisfy all the customers' requirements yet as IP technologies still have drawbacks.

The disadvantages of VoIP services:

VoIP phones depend on wall power. Whether you use VoIP software installed on your PC or VoIP phones you are dependent on power while your conventional phone relies on the power supplied by a separate line from the central office. If your power goes off you still can use your ordinary phone but not the VoIP phone (unless it is powered by batteries).

Many other systems you may use in your home are integrated with conventional phone lines – digital video recorders, subscription TV services, home security systems and others still can't be integrated with VoIP.

Calling 911 with VoIP can cause problems. An emergency 911 call from an ordinary phone is received by the 911 operator and your current location can be easily identified. But IP address used by VoIP can't tell where you are situated and the central 911 office does not know where to route your call and what is the nearest emergency service station to you. VoIP providers and developers try to solve this issue but including geographic information into IP address may deprive VoIP of its other benefits.

VoIP uses Internet connection therefore all the factors that affect the quality of your internet are to affect the quality of your conversation. Latencies, jitters, packet losses may distort or even frustrate your session.

As any other information stored on your computer and transmitted through Internet Protocol VoIP is susceptible to viruses and hacking.

Much depends on the processor your computer uses and other requirements. If you run several programs simultaneously your VoIP phone call may be distorted. The program may either slow down or even crash in the middle of an important conversation.

VoIP providers and developers are now working to solve these issues and optimize the benefits of VoIP [6].

## VIII.   IP TELEPHONY MARKET

As VoIP continues to improve in quality and security becomes less of a threat when partnered with robust applications, the technology is gaining traction. Enterprises are realizing that the benefits of IP telephony exceed basic cost reduction.

Companies have also come to appreciate the fact that merging voice and data into a single network provides an enhanced communication experience. As a result, these organizations are increasingly interested in the convergence that IP communication grants and the integration of multiple applications in a single interface or device.

According to new analysis from Frost & Sullivan, North American Enterprise IP Telephony End-Point Markets, 2006 showed market earned revenues of $1.02 billion. This market is estimated to reach $2.79 billion in 2011 [8].

Analyst firm Dell'Oro Group sees the carrier IP Telephony market breaking out of its current slump in 2010, according to a recent report. While the group said the carrier IP telephony market in 2009 will be down around 14 percent from 2008, Dell'Oro expects the market to rebound to $4 billion by 2013 [9].

By the end of 2008, the Asia-Pac enterprise telephony market is predicted to grow by 9.4 percent (year-on-year) to reach revenues of close to $2.98 billion, with IP telephony estimated to account for 59.2 percent ($1.76 billion) of this total.

According to Frost & Sullivan, the main motivator behind IP deployments is the need to bridge present-day enterprise communication needs through the use of next-generation applications, which enables convenience, cost savings and enhanced productivity [10].

Converged communications are becoming increasingly important in the enterprise. As such, those operating in this space must be proactive by anticipating trends and demands and offering products and services that address the challenges that organizations face in trying to achieve seamless integration of both data and voice. By doing so, a win-win situation is created, helping to drive growth for both vendors and targeted organizations [8].

## IX.   CONCLUSION

VoIP is really "Everything over IP". VoIP and 802.11 technologies are relatively young; many businesses are unwilling to commit a critical communications infrastructure to them until they have proven themselves.

Many businesses are moving to wireless VoIP. They tend to be in highly mobile industries. Major regulatory issues raised by VoIP. The biggest challenge towards quicker and larger-scale uptake of IP telephony is the issue of legacy equipment. Concerns relating to VoIP really relate to convergence generally.